\documentclass[pre,superscriptaddress,showpacs,reprint]{revtex4-1}
\usepackage[latin1]{inputenc}
\usepackage[dvips]{graphicx}
\usepackage{verbatim}
\usepackage{amsmath}
\usepackage{amsfonts}
\usepackage{amssymb}
\usepackage{latexsym}

\usepackage{xcolor}

\newcommand{\beq}{\begin{equation}}
\newcommand{\eeq}{\end{equation}}

\begin{document}

\title{Preferential attachment and power-law degree distributions in heterogeneous multilayer hypergraphs}

\author{Francesco Di Lauro, Luca Ferretti} 
\affiliation{Pandemic Sciences Institute and Big Data Institute, Nuffield Department of Medicine, University of Oxford, Oxford, United Kingdom,}
\email{Email: luca.ferretti@gmail.com; luca.ferretti@bdi.ox.ac.uk}

\begin{abstract}
We include complex connectivity structures and heterogeneity in models of multilayer networks or multilayer hypergraphs growing by preferential attachment. We consider the most generic connectivity structure, where the probability of acquiring a new hyperlink depends linearly on the vector of hyperdegrees of the node across all layers, as well as on the layer of the new hyperlink and the features of both linked nodes. We derive the consistency conditions that imply a power-law hyperdegree distribution for each class of nodes within each layer and of any order. For generic connectivity structures, we predict that the exponent of the power-law distribution is universal for all layers and all orders of hyperlinks, and it depends exclusively on the type of node.
\end{abstract}


\maketitle

\section{Introduction}

Scale-free degree distributions are one of the most interesting features of complex networks \cite{caldarelli_scale-free_2007}. Preferential attachment is one of the most well-understood general explanations for this feature \cite{barabasi_emergence_1999}, and a plethora of variations around models of growing networks with preferential attachment have been published for over two decades. 

A feature that is bound to be present in almost any realistic model is the heterogeneity between nodes and links. Although systems in which all nodes and links are identical to each other can be found in some physical and technological networks, most social and biological networks are composed of different types of node features, connected by different kinds of links or interactions. Preferential attachment models with heterogeneous nodes have been already studied in the past ~\cite{BianconiBarabasi,bianconi_bose-einstein_2001, ferretti_preferential_2011, ferretti_duality_2014,rosengren_multi-type_2018}, focusing on heterogeneities such as different node fitness \cite{BianconiBarabasi,dorogovtsev_structure_2000}, different spatial location \cite{ferretti_preferential_2011}, and even completely arbitrary heterogeneous node features \cite{ferretti_features_2012}. Heterogeneous links can be framed in the context of multi-layer networks \cite{de_domenico_mathematical_2013,boccaletti_structure_2014}, and more specifically multiplex networks, as an heterogeneous set of layers, with each layer characterised by a specific type of link. Several papers derived results on preferential attachment in multi-layer networks \cite{rosengren_multi-type_2018,kim_coevolution_2013,momeni_growing_2015,backhausz_asymptotic_2019}, showing how these models create a very strong correlation between the layer-specific degree of each node across different layers \cite{kim_coevolution_2013}.

Most of these preferential attachment models can also be easily generalised to hypergraphs, or higher-order networks \cite{bianconi_higher-order_2021} by extending the classical preferential attachment rule to hypergraphs \cite{guo_non-uniform_2016, zhou_emergence_2020, guo_emergence_2023}.

In this work, we describe a wide class of general preferential attachment multi-layer hypergraph models with arbitrary connectivity structure, arbitrary heterogeneity in both nodes and layers of hyperlinks. We derive the general formalism to compute the tail of the degree distribution for these models, showing how the problem reduces to a matrix generalisation of the consistency equations in \cite{BianconiBarabasi, ferretti_features_2012}. We derive the exponent of the power-law degree distribution for special cases. Finally, to illustrate the power of this formalism, we study the degree distribution in a series of examples, including a Bianconi-Barab\'asi fitness model on hypergraphs and a network model where each node can be global or local, i.e. present in all layers or just in a single one. 

\section{The model}

The model is built on the classical structure of the Bar\'abasi-Albert model \cite{barabasi_emergence_1999} and its extensions with features, i.e. different types of nodes \cite{ferretti_features_2012} and some simple models with multiple layers \cite{rosengren_multi-type_2018,kim_coevolution_2013,momeni_growing_2015,backhausz_asymptotic_2019}. We extend these models to hypergraphs with generic heterogeneity in interactions across nodes and layers (and orders). We formulate the model in terms of multiplex networks, in which each node exists across all layers and each link belongs to a single layer; however, the class of models we describe includes any type of multilayer networks.

Note that generalising preferential attachment to hypergraphs presents an ambiguity, since the standard preferential attachment (based on node degree) can be generalised either by considering the hyperdegree of each node (link-based preferential attachment) or by considering the number of his neighbour (neighbour-based preferential attachment). In this work, we consider both possibilities.

\paragraph*{Notation:} types of nodes are indicated by Greek letters; the set of all types of nodes is denoted by $\mathcal{N}$. Network layers are indicated by Latin uppercase letters; the set of all layers is denoted by $\mathcal{L}$. Hyperlink orders are denoted by Greek uppercase letters; the set of all hyperlink orders is denoted by $\mathbf{\Theta}$. The indexes of individual nodes are indicated by lowercase letters; the type of node $i$ is denoted by $\tau(i)$. Finally, $k_{i,\Lambda,A}$ denotes the hyperdegree (number of hyperlinks of order $\Lambda$) of node $i$ in layer $A$.

\paragraph*{Growth:} the growth of the network or hypergraph follows these steps:

\begin{itemize}
\item start from a small random hypergraph;

\item at every time step, extract a type $\alpha$ with probability $p_\alpha$ and add a new node of type $\alpha$. The node is born with $m_{\alpha,\Lambda,A}$ hyperlinks of order $\Lambda$ in layer $A$;

\item the probability that a link in layer $A$ is connected from the new node to an existing node $i$ is 
\beq
\Pi_{i,\Omega,A}=\frac{\sum_{\Lambda,B}\sigma_{\tau(i);\alpha,\Omega,A;\Lambda,B}w(\Lambda)k_{i,\Lambda, B}}{\sum_j\sum_{ \Lambda,B}\sigma_{\tau(i);\alpha,\Omega,A;\Lambda,B}w(\Lambda)k_{j,\Lambda, B}}
\eeq
where all the entries of the hypermatrix $\sigma_{\beta;\alpha,\Omega,A;\Lambda,B}$ are non-negative, and $w(\Lambda)$ is either $1$, for link-based preferential attachment, or $\Omega -1$ for neighbour-based. 
This hypermatrix, together with the initial connectivity $m_{\alpha,\Lambda,A}$, defines the connectivity structure of the network.
\end{itemize}

\section{Degree distribution}

\subsection{Evolution of a node}

The evolution of node degrees can be described in the continuum approximation \cite{barabasi_emergence_1999} by the equation:
\beq
\frac{dk_{i,\Omega,A}}{dt}=\sum_\alpha p_\alpha m_{\alpha,\Omega,A} \left(\Omega-1\right) \Pi_{i,\Omega,A}. 
\eeq
For large networks, this equation is solvable using the ansatz 
\beq
\sum_j\sum_{\Lambda,B}\sigma_{\tau(j);\alpha,\Omega,A;\Lambda,B} w(\Lambda)k_{j,\Lambda,B}\simeq C_{\alpha,\Omega,A}t+o(t),\label{eq:ansatz}
\eeq
where the coefficients $C_{\alpha, \Omega,A}\geq 0$ will be derived in the next section. Hence, ignoring small corrections in $t$, the equation for the degree of node $i$ is
\beq
\frac{dk_{i,\Omega,A}}{dt}=\sum_\alpha p_\alpha m_{\alpha,\Omega,A} (\Omega-1)
\sum_{B,\Lambda}\frac{\sigma_{\tau(i);\alpha,\Omega,A;\Lambda,B}w(\Lambda)}{C_{\alpha,\Omega,A}}\frac{k_{i,\Lambda, B}}{t}.\label{eq:ksimpl}
\eeq
We define the quality matrix $Q_\tau$ in layer space:
\beq
Q_{\tau;\Omega,A;\Lambda,B}=\sum_\alpha p_\alpha m_{\alpha,\Omega,A}\left(\Omega-1\right)\frac{\sigma_{\tau;\alpha,\Omega,A;\Lambda,B}}{C_{\alpha,\Omega,A}},\label{eq:qdef}
\eeq
Note that all entries of $Q_\tau$ in this base are non-negative. Then we can rewrite equation (\ref{eq:ksimpl}) in matrix-vector notation in layer space to obtain
\beq
\frac{d\mathbf{k}_{i}}{dt}=Q_{\tau(i)}\frac{\mathbf{k}_{i}}{t},
\eeq
which is solved analytically by matrix exponentiation:
\beq
\mathbf{k}_i(t)=e^{Q_{\tau(i)}\log(t/t_0)}\mathbf{k}_i(t_0)=e^{Q_{\tau(i)}\log(t/t_0)}\mathbf{m}_{\tau(i)},
\eeq
where $t_0$ is the birth time of the node.

We denote by $q^{max}_\tau$ the leading eigenvalue of $Q_{\tau}$ and by $\chi^{max}_{\tau}$ the corresponding eigenvector. In the generic case, we assume that $Q_{\tau}$ is diagonalisable, $\chi^{max}_{\tau}$ is unique and all its components are positive, and that the coefficient $\mu_\tau$ of $\chi^{max}_{\tau}$ in the eigenvector decomposition of $m_\tau$ is positive; we will relax these assumptions in a later section. Under these assumptions, the leading behaviour of the node degree is 
\beq
k_{i,\Omega,A}(t)\approx (t/t_0)^{q^{max}_{\tau(i)}} \mu_{\tau(i)} \chi^{max}_{\tau(i),\Omega,A}.\label{eqk}
\eeq

Note that a consistency requirement is $q^{max}_\tau\leq 1$, in order to avoid an impossible superlinear growth of the nodes \cite{BianconiBarabasi,ferretti_preferential_2011,ferretti_features_2012}. 

\subsection{Consistency equations}

The positive coefficients $C_{\alpha,A,\Omega}$ can be implicitly determined by a set of consistency conditions. For a single layer, the equations reduce to the ones in \cite{ferretti_features_2012} for heterogeneous models with preferential attachment and features. Here we derive the general consistency equations for multilayer networks and hypergraphs. These equations are exact even if derived under the continuum approximation (see appendix C in \cite{ferretti_features_2012}).

The overall hyperdegree $K_{\alpha,\Omega,A}=\sum_{\tau(i)=\alpha}k_{i,\Omega,A}(t)$ at order $\Omega$, layer $A$, of all nodes of type $\alpha$ is
\beq
K_{\alpha,\Omega,A}(t)=p_\alpha\int_1^t dt_0\sum_{ \Lambda, B} (e^{Q_{\alpha}\log(t/t_0)})_{\{\Omega,A\},\{\Lambda,B\}}m_{\alpha,\Lambda,B}.
\eeq
The matrix integral can be easily solved in vector notation to obtain
\beq
\mathbf{K}_{\alpha}(t)=tp_\alpha\left[(1-Q_\alpha)^{-1}\left(1-e^{-(1-Q_\alpha)\log(t)}\right)\right]\mathbf{m}_{\alpha}.
\eeq
If all eigenvalues of $1-Q_\alpha$ are strictly positive (i.e. outside of the condensation phase), for large times we find
\beq
\mathbf{K}_{\alpha}(t)=tp_\alpha(1-Q_\alpha)^{-1}\mathbf{m}_{\alpha}.
\eeq
Inserting this result into the ansatz (\ref{eq:ansatz}), the consistency equations can then be written as
\begin{align}
C_{\alpha,\Omega,A}= &\sum_\beta\sum_{\Lambda,B}p_\beta\sigma_{\beta;\alpha,\Omega,A;\Lambda,B} w(\Lambda)\nonumber\\
&\sum_{\Gamma,D}\left[(1-Q_\beta)^{-1}\right]_{\{\Lambda,B\},\{\Gamma,D\}}m_{\beta,\Gamma,D},
\end{align}
which determine the coefficients $C_{\alpha,\Omega,A}$ together with the equations (\ref{eq:qdef}).

If there is a finite number of types of nodes, that is, if $|\mathcal{N}|<+\infty$, the above consistency equations should always admit a solution such that all eigenvalues of $1-Q_\alpha$ are strictly positive. 

On the other hand, if $|\mathcal{N}|=+\infty$, solutions may not exist at all, either because some of the sums could diverge (heterogeneity-driven attachment) \cite{ferretti_features_2012} or because $1-Q_\alpha$ contains null eigenvalues for some types $\alpha$. The existence of null eigenvalues of $1-Q_\alpha$ implies that $q^{max}_\alpha=1$, that is, at least some nodes of type $\alpha$ tend to grow almost linearly in time. This corresponds to a different phase, i.e. Bose-Einstein condensation of links on a small number of nodes \cite{BianconiBarabasi,ferretti_features_2012,Ferretti2014}. In the following, we will consider only networks outside the condensation phase, i.e. $q^{max}_\alpha<1$ for all types $\alpha$.

\subsection{Multilayer degree distribution}

Generic connectivity structures correspond to generic matrices $Q_\tau$ that satisfy the following assumptions: (i) $Q_\tau$ is diagonalisable, (ii) its highest eigenvalue $q^{max}_\tau$ has geometric multiplicity 1, and (iii) all the components of the corresponding eigenvector are positive. For a generic structure, we also assume that (iv) the decomposition of $m_\tau$ in terms of eigenvectors of $Q_\tau$ has first coefficient $\mu_\tau>0$.

 Given the evolution of node degree $k_{i,\Omega,A}$ described in equation (\ref{eqk}), the degree distributions for nodes of type $\alpha$ can be easily obtained from the continuum approximation
 \beq
 p(k_{\alpha,\Omega,A})\sim k_{\alpha,\Omega,A}^{-\left[1+1/q^{max}_{\alpha}\right]}.
 \eeq
Interestingly enough, the exponent of this power-law is \emph{universal}, i.e. it is the same for all layers! For hypergraphs, it is also the same for all orders. The distribution changes with node type $\alpha$ since $q^{max}_{\alpha}$ depends on node type, but does not change with respect to the layers or orders. 

As already observed in the literature on multilayer networks, the equations for the evolution of node degrees also imply a strong correlation between degrees $k_{\alpha,A},k_{\alpha,B}$ from the same types of node $\alpha$ but different layers $A,B$~\cite{kim_coevolution_2013}. The same is true for different orders of hyperlinks. In fact, if the evolution of node degree would follow the continuous deterministic process, the correlation  between the hyperdegree of nodes of a given type $\tau(i)=\alpha$ across different layers or orders would be close to $\mathrm{Cor}\left[k_{i,\Omega,A},k_{i,\Lambda,B}\right]\approx 1$ for older nodes. This corresponds to the linear relation between degrees at different layers or orders
\beq
\frac{k_{i,\Omega,A}}{k_{i,\Lambda,B}}\approx \frac{\chi^{max}_{\alpha,\Omega,A}}{\chi^{max}_{\alpha,\Lambda,B}}.
\eeq



\subsection{A simple network example with fitness and cross-layer preferential attachment}

As a simple example, consider an extension of the classical fitness model by Bianconi and Barab\'asi \cite{BianconiBarabasi} to networks with two asymmetric layers. The first layer grows according to the Bianconi-Barab\'asi fitness model with fitness $\eta$, while the second evolves by preferential attachment based on the node degree in the first layer weighted by the fitness $\eta'$. Node classes are defined by their joint fitness $\tau=(\eta,\eta')$ and are randomly extracted from the fitness distribution with density $p_\tau=\rho(\eta,\eta')d\eta d\eta'$. Hence, the $2\times 2$ layer connectivity hypermatrix $\sigma_{\tau;\alpha,\cdot;\cdot}=\begin{pmatrix} \eta&0\\ \eta'&0 \end{pmatrix}$ is independent on $\alpha$. This implies also that $C_{\alpha,\cdot}=(C,C')$ is independent on $\alpha$. All nodes begin with initial connectivity $m_\tau=(m,m')$ respectively. 

The quality matrix is
\beq
Q_{(\eta,\eta')}=\begin{pmatrix} m\eta/C&0\\ m'\eta'/C'&0 \end{pmatrix},
\eeq
which has a null eigenvalue and a maximum eigenvalue $q^{max}_{\eta,\eta'}=m\eta/C$. The consistency equations for $C,C'$ are
\begin{align}
C=&\int d\eta \rho(\eta) \frac{m\eta}{1-m\eta/C}, \\
C'=&\int d\eta d\eta' \rho(\eta,\eta') \frac{m\eta'}{1-m\eta/C},
\end{align}
where $\rho(\eta)=\int d\eta' \rho(\eta,\eta')$ is the marginal distribution for $\eta$. 

 The eigenvector corresponding to the leading eigenvalue is $(m\eta/C,m'\eta'/C')$, hence 
\beq
\begin{pmatrix} k\\ k' \end{pmatrix}\rightarrow \begin{pmatrix} m\eta/C\\ m'\eta'/C' \end{pmatrix}\left({t}/{t_0}\right)^{\frac{m\eta}{C}}.
\eeq
Hence, node growth and degree distribution are controlled by the fitness $\eta$ of the first layer only. The degree distribution is $p(k)\sim k^{-\left(1+\frac{C}{m\eta}\right)}$ for both layers, in agreement with the general form. The first layer behaves precisely like the classical Bianconi-Barab\'asi model, while the fitnesses $\eta'$ have only a multiplicative effect on the asymptotic growth of the degrees in the second layer.

\subsection{Constraints on connectivity}
The discussion above is valid for a complex connectivity structure for which the assumptions above (i-iv) hold. For a generic matrix $Q_\tau$ or vector $m_\tau$ with uncorrelated positive entries, these assumptions are realised. However, there could be symmetries or constraints in the connections that violate these assumptions. For example, several layers could share the same properties, or some node types could not have links in some layers (e.g. they do not exist in those layers) and so on. In this section, we discuss the special cases where these assumptions are not met.

If assumption (i) does not hold, the Jordan canonical form $Q_\tau^J$ of $Q_\tau$ is not diagonal. Let us consider the diagonal block of $Q_\tau^J$ corresponding to the highest eigenvalue. The block has matrix form $q^{max}_\tau\delta_{i,j}+\delta_{i+1,j}$ and the leading term of its exponential is 
\beq
k_{\tau,\Omega,A}(t)\propto \log(t/t_0)^{d_\tau-1}\cdot (t/t_0)^{q^{max}_\tau} \mu_\tau \chi^{max}_{\tau,\Omega,A},
\eeq 
where $d_\tau$ is the dimension of this block and $\chi^{max}$ is the proper eigenvector in the block. Hence, asymptotically the only difference is a logarithmic correction to the power growth of node degree.

If (i) is true but (ii) is not, then there are multiple eigenvectors corresponding to $q^{max}_\tau$. The degree distribution for nodes of type $\tau$ has the same exponent $1+1/q^{max}_\tau$, but the asymptotic proportion of links in each layer could depends on stochastic effects. The average distribution is proportional to $\sum_k\mu_{\tau,(k)}\chi^{max}_{(k)\tau,\Omega,A}$ where $\chi^{max}_{(k)\tau}$ are the multiple eigenvectors of $q^{max}_\tau$. 

If (i) and (ii) are valid but (iii) does not hold, then $\chi^{max}_{\tau}$ has some null components. This is possible only if $Q_\tau$ has a block-upper-diagonal form for some permutation of layers, i.e. if there is a subset of layers such that their growth depends only on their links and not on links in other layers. In this case, we can write the full decomposition of $m_{\tau,\Omega,A}=\sum_k\mu_{\tau,k}\chi_{\tau,k,\Omega,A}$ with eigenvectors $\chi_{\tau,k,\Omega,A}$ ordered by decreasing eigenvalues $q_{\tau,k}$. For each order/layer pair $\Omega,A$, we consider the minimum index $j_\tau(\Omega,A)$ such that $\chi_{\tau,j_\tau(\Omega,A),\Omega,A}\neq 0$. Without loss of generality, we can choose $\chi_{\tau,j_\tau(\Omega,A),\Omega,A}> 0$. Then the degree for layer $A$ grows as
\beq
k_{i,\Omega,A}(t)\approx (t/t_0)^{q_{\tau(i),j_{\tau(i)}(\Omega,A)}} \mu_{\tau(i),j_{\tau(i)}(\Omega,A)} \chi_{\tau(i),j_{\tau(i)}(\Omega,A),\Omega,A}
\eeq
and the corresponding hyperdegree distribution has exponent $1+1/q_{\tau,j_\tau(\Omega,A)}$ that now depends on the order of the links and the layer as well.

Finally, it is possible (i), (ii) and (iii) are true but (iv) is not. In that case, stochastic effects cause $k_i(t)$ to acquire almost surely a component $\chi^{max}_{\tau(i)}$. Hence, equation (\ref{eqk}) is valid but with $\mu_\tau$ replaced by a random coefficient that depends on the node.


\subsection{Special cases}
\subsubsection{Networks}

\paragraph*{Networks with identical nodes:} If there is a single type of nodes, the model is determined by the connectivity matrix $\sigma_{A;B}$ and the initial degree $m_A$. In this case, since $q^{max}$ does not depend on the layer, there is a single $q^{max}$ for all nodes and layers. The equations can be solved as
\begin{align}
Q_{A,B} &= \frac{\sigma_{A;B}m_A}{2\sum_{D}\sigma_{A;D}m_D}, \\
C_A &= 2\sum_{B}\sigma_{A;B}m_B.
\end{align}

Note that $2Q_{A,B}m_B/m_A$ is a the transpose of a transition matrix, hence it has a single maximum eigenvalue equal to 1, corresponding to the right eigenvector $\mathbf{1}=(1\ldots 1)$. This implies that, asymptotically, the balance between addition of nodes and links causes the quality to reduce to the Barabasi-Albert one $q^{max}=1/2$. The corresponding eigenvector is $\chi_A=m_A$, and it is the only one if $Q_{A,B}$ is irreducible. 

Hence, the tail of the degree distributions for all layers behave like $p(k)\sim k^ {-3}$ as in the classical Barabasi-Albert model, with $k_{i,A}/k_{i,B}\rightarrow m_A/m_B$ up to small stochastic effects. Note that the amount of links in each layer is given by $2m_At$ and differs among layers. 

\paragraph*{Networks with homogeneous node spaces:} The previous example belongs to a large class of models whose connectivity properties are symmetric with respect some exchange of node types. In these models, each type of node can be mapped into every other type by a symmetry transformation belonging to a subgroup of the permutations of $\mathcal{N}$, preserving both $p_\alpha$, $m_{\alpha,A}$ and $\sigma_{\beta;\alpha,A;B}$, i.e. the node space is homogeneous \cite{ferretti_preferential_2011,ferretti_features_2012}. This implies $p_\alpha=1/|\mathcal{N}|$ and $m_{\alpha,A}=m_A$. Note that we assume this symmetry property applies only to nodes, while layers can be extremely asymmetric -- as in the previous case. 

The symmetry of this class of models implies that $q^{max}_\tau$ does not actually depend on $\tau$, hence the universal quality $q^{max}=1/2$ does not depend on node or layer and the distribution for each layer behaves as $p(k)\sim k^ {-3}$ as in the previous case, with a relative degree between layers $k_{i,A}/k_{i,B}\rightarrow m_A/m_B$ as well. 

\paragraph*{Networks with homogeneous layer sets:} If the reverse is true, i.e. layers are symmetric while nodes are heterogeneous, then the equations are greatly simplified and reduce to a set similar to those of heterogeneous models \cite{ferretti_features_2012} for a set of independent layers. 

The permutation symmetry implies that $C_{\alpha,A}=C_\alpha$ and $m_{\alpha,A}=m_\alpha$ do not depend on the layer. Further, if $v_\tau$ is an eigenvector of $Q_\tau$, then all transformed vectors $T(v_\tau)$ are eigenvectors corresponding to the same eigenvalue, and therefore $\mathbf{1}\propto\sum_TT(v_\tau)$ is an eigenvector for the same eigenvalue as well. This implies that there is only one eigenvalue $q_\tau$ for the whole $Q_\tau$ matrix. 

In turn, this implies a power-law distribution for nodes of type $\tau$ with the same tail 
\beq
p(k_\tau)\sim k_\tau^{-[1+1/q_\tau]},
\eeq
where $q_\tau$ is defined by the same consistency equations for all layers
\begin{align}
q_\tau &= \sum_\alpha\frac{p_\alpha\sigma_{\tau;\alpha}m_\alpha}{C_\alpha}, \\
C_\alpha &= \sum_\beta \frac{p_\beta\sigma_{\beta;\alpha}m_\beta}{1-q_\beta},
\end{align}
 defining $\sigma_{\tau;\alpha}=\sum_{B}\sigma_{\tau;\alpha,A;B}$. These equations closely resemble the ones in section IV.A of \cite{ferretti_features_2012} for a single layer.

\subsubsection{Hypergraphs}

\paragraph*{Hypergraphs with identical nodes:}

In an hypergraph model with a single type of node, we also write explicit equations for both $q_{\text{max}}$ and $C$ (see section \ref{fitnesshg} for a derivation):
\begin{align}
    C &= \sum_{\Lambda} w(\Lambda) \Lambda m_{\Lambda}, \nonumber \\
    q_{\text{max}} &= \frac{  \sum_{\Lambda} m_\Lambda w(\Lambda) \left(\Lambda -1\right) }
    {\sum_{\Lambda} m_{\Lambda} w(\Lambda) \Lambda }. \label{exact_eq_qmax_hyper}
\end{align}

For classic networks ($\Lambda =2)$, we get $q_{\text{max}} = \frac{1}{2}$ and the classical exponent $1+1/q_{\text{max}}=3$. In the general case, the exponent of the power-law tail of the degree distribution differs if the model is built on link-based or neighbour-based preferential attachment. However, if higher orders dominate, this quantity then $q_{\text{max}} \to \frac{1}{1-\Omega_{\text{max}}}$ in both cases. The exponent of the degree distribution tends to 
\beq
1 + \frac{1}{q_{\text{max}}} \to \frac{2\Omega_{\text{max}} -1}{\Omega_{\text{max}} -1}, \label{qmax_higherorder}
\eeq
which is between 2 and 3.

\paragraph*{Hypergraphs with homogeneous node spaces:}
In these models, each type of
node can be mapped into every other type by a symmetry transformation which preserves $p_\alpha$, $m_{\alpha,\Omega,A}$ and $\sigma_{\beta;\alpha,\Omega,A;\Lambda,B}$, i.e. the node space is homogeneous. In this case, because of the symmetry among all nodes, the equations (\ref{exact_eq_qmax_hyper}) and  (\ref{qmax_higherorder}) for the tail of the distribution are still valid.


\subsection{More complex examples}
\subsubsection{Fitness model on a hypergraph}\label{fitnesshg}
To highlight the differences between a network and a higher-order network, we present the results for a fitness model a la Barab\'asi-Bianconi but with an arbitrary number of orders,  constant fitness independent of the order $\sigma_{\eta,\Omega, \Lambda} = \eta$, and only one layer. Similarly as before, node classes are identified by their fitness $\eta$. The Q matrix of this model is
\beq
Q_{\eta, \Omega, \Lambda} = \sum_{\alpha} p_\alpha m_\Omega \left( \Omega -1 \right)\frac{\eta}{C_{\alpha,\Omega}} w(\Lambda);
\eeq
the consistency equation for $C$ becomes
\beq
C = \sum_{\beta} \sum_{\Lambda} p_\beta \beta w(\Lambda) \sum_{\Gamma} \left[\left(1 - Q_\beta\right)^{-1} \right]_{\Lambda, \Gamma}m_{\Gamma},
\eeq
which is constant with respect to $\alpha$ and $\Omega$. This makes $Q$ a projector with eigenvector $m_\Omega (\Omega-1)$. This means that we can implicitly compute $C$, by Taylor expansion:
\begin{align}
C &= \sum_{\beta} \sum_{\Lambda} p_\beta \beta w(\Lambda) \sum_{\Gamma} w(\Lambda)  \left(\mathbf{1} + \sum_{k\geq 1} Q^k \right) m_\Gamma \nonumber \\
&=\sum_{\beta} p_\beta \beta \sum_{\Gamma, \Lambda} w(\Lambda) \left( \delta_{\Gamma, \Lambda} + Q_{\beta, \Lambda, \Gamma} \frac{1}{1-h(\beta)}\right) m_\Gamma \nonumber \\
&= \sum_{\beta} p_\beta \beta \left(1-\frac{\beta \sum_\Lambda w(\Lambda) (\Lambda-1) m_\Lambda}{C}\right)^{-1}, 
\end{align}
where we used that $Q$ behaves like a projector, so 
\begin{align}
Q^k_{\eta,\Omega,\Lambda} &= Q_{\eta,\Omega,\Lambda} \left[ \frac{\eta}{C} \sum_\Gamma m_\Gamma \left( \Gamma -1\right) w(\Gamma)\right]^{k-1} \nonumber \\
&\equiv Q_{\eta,\Omega,\Lambda} \, \left[h\left(\eta\right) \right]^{k-1}.
\end{align}

We note that $h(\eta)=q_{\text{max}}$ is the maximum eigenvalue of $Q$.

Now we can highlight the differences between the hypergraph fitness model and the Barab\'asi-Bianconi equations [cite]. The eigenvalue for the latter is $q_{BB} = \frac{\eta}{C_{BB}}$,  with $C_{BB} = \sum_{\beta} p_\beta \beta \frac{1}{1-\frac{\beta}{C_{BB}}}$. Rewriting the equation for $q_{\text{max}}$ in the same form as $q_{BB}$, we get $$
q_{\text{max}} = \frac{\eta}{C/\sum_\Gamma m_\Gamma (\Gamma-1) w(\Omega)} \equiv \frac{\eta}{\tilde{C}},
$$
where the consistency equation for $\tilde{C}$ becomes
$$
\tilde{C} = \frac{\sum_{\Gamma} m_\Gamma w(\Gamma)}{\sum_\Gamma m_\Gamma (\Gamma -1) w(\Gamma)} \sum_\beta p_\beta \beta \frac{1}{1-\tilde{C}},
$$
from which we conclude that, if there are higher orders $\Omega>2$, then $\tilde{C} < C$. This means that the power-law exponents for the hypergraph model are smaller, and the phase transition to a Bose-Einstein condensate happens earlier than in the Barab\'asi-Bianconi model, although this could be delayed because individual nodes have to share hyperlinks with $\Omega-1$ nodes (see \cite{Ferretti2014} for a model where condensation is delayed for similar reasons).

\subsubsection{A multilayer network with global and local nodes}
In this section, we study a simple but interesting multilayer network model where each node can be global or local, i.e. present in all layers or just in a single one. This model can be thought of as a hybrid between multiplex and interconnected networks.

We consider a model with $n$ layers (denoted by $1\ldots n$) and $n+1$ types of nodes
. All layers grow according to the classical preferential attachment based on the overall node degree computed across all layers. 
Nodes of the $l(D)$ type 
exist only in the $D$th layers, i.e. they can share links only in the $D$th layer, while nodes of type $0$ exist and can connect across all layers. We refer to these nodes as local and global nodes, respectively. The fraction of global nodes is $p_g$, while local nodes are equally distributed across all layers, so that each layer contains s $p_{l(D)}=p_l=\frac{1-p_g}{n}$ of local nodes. The initial number of connections across all layers is $m_g$, for global nodes -- equally distributed across all layers -- and $m_{l}$, for local nodes, respectively. The hypermatrix for preferential attachment is therefore $\sigma_{g;\cdot,A;B}=\sigma_g$ and  $\sigma_{l(D);\cdot,A;B}=\sigma_l\delta_{A,D}\delta_{B,D}$.

This model is invariant under permutations acting jointly on the set of layers and the local nodes. Hence, although neither the nodes nor the layer space are homogeneous, the model can be solved by an approach similar to that outlined for homogeneous sets.

Using the symmetries of the model and the properties of the connectivity hypermatrix, the general shape of the quality matrices is
\begin{align}
Q_{g,A,B}&=\frac{q_g}{n},\\
Q_{l(D),A,B}&=q_l\delta_{A,D} \delta_{B,D},
\end{align}
The first derives from the symmetry under layer permutations and from the fact that the connectivity hypermatrix $\sigma_{g;\cdot,A;B}$ for  global nodes does not depend on $B$ by construction; the second is obvious since local nodes in set $l(D)$ can have links only in layer $D$ and all layers are identical. 

Surprisingly enough, both normalisations $C_{g,A}=C$ and $C_{l(D),A}=C\delta_{A,D}$ can be expressed in terms of a single quantity $C$ satisfying the consistency condition 
\beq
1=\frac{p_g\sigma_gm_g}{C/n-(p_gm_g+p_lm_l)\sigma_g}+\frac{p_l\sigma_lm_l}{C-(p_gm_g+p_lm_l)\sigma_l}.
\eeq
The final qualities can be expressed in terms of $C$ as
\begin{align}
q_g&=n\frac{p_gm_g+p_lm_l}{C}\sigma_g,\\
q_l&=\frac{p_gm_g+p_lm_l}{C}\sigma_l.
\end{align}

To understand the result, it is useful to consider two extreme cases: $n=1$ and $n\gg 1$. 

For a single layer $n=1$, the qualities are the same for local and global nodes up to the ratio of their ``attractiveness'' $q_g=\frac{\sigma_g}{\sigma_l}q_l$. 

In the case of large $n\gg 1$, since $p_l,m_g\sim O(1/n)$, then the normalisations converge to $C\rightarrow n(2p_gm_g+p_lm_l)\sigma_g$. This implies the qualities
\begin{align}
q_g&\rightarrow \frac{p_gm_g+p_lm_l}{2p_gm_g+p_lm_l},\\
q_l&\rightarrow \frac{1}{n}\frac{p_gm_g+p_lm_l}{2p_gm_g+p_lm_l}\frac{\sigma_l}{\sigma_g},
\end{align}
that is, the global nodes follow a growth dynamics with a modified exponent $q_g=\frac{p_gm_g+p_lm_l}{2p_gm_g+p_lm_l}$ and therefore a power-law  degree distribution with exponent between 2 and 3:
\beq
p(k_g)\sim k_g^{-\frac{3p_gm_g+2p_lm_l}{p_gm_g+p_lm_l}},
\eeq
while the local ones grow much more slowly and have a negligible power-law tail in their degree distribution.

Hence, for a low number of layers, the attractiveness of global versus local nodes is the main determinant of the shape of the power-law degree distributions of both classes of nodes. On the other hand, for a large number of layers, local nodes become less and less relevant, while the power-law degree distribution of global nodes depends only on the fraction and initial connectivity of both classes of nodes.

\section{Discussion}

In this paper we presented the first complete analysis of degree distributions in multilayer higher-order networks with preferential attachment and arbitrary node and layer structures. The hyperlink structure is also quite general, albeit it is restricted to hyperlinks that cannot grow in their order while the network grows. Models with hyperlink that can increase their order result in a different structure \cite{owada_degree_2024}.

Note that while these models are naturally described in terms of multiplex networks, the formulation is general enough to apply to any multilayer network, including interconnected networks. In fact, interconnected networks can be simply described by a model in which a ``layer" actually corresponds to a type of node $\alpha\in\mathcal{N}$. The space of possible classes of inter- and intra-``layer" links would then be $\mathcal{L}=\mathcal{N}\times\mathcal{N}$, with the diagonal components of the Cartesian product describing the intra-``layer" connectivity and the off-diagonal component the inter-"layer" ones. The parameters of the model should be further constrained as $m_{\alpha,\Lambda,\{\beta,\gamma\}}=0$ if $\alpha\neq\beta$, and $\sigma_{\beta;\alpha,\Omega,\{\gamma,\delta\};\Lambda,\{\epsilon,\phi\}}=0$ if $\beta\neq\delta$ (or $\alpha\neq\gamma$).

The directed version of the models we discuss here -- i.e. with preferential attachment proportional to the incoming degree $k_{in}$ plus a constant -- should show a similar asymptotic dynamics as the models discussed here, since $k_{out}$ is fixed for each type of nodes, hence $k\sim k_{in}$ for older nodes, and the constant does not play any role \cite{dorogovtsev_structure_2000,krapivsky_connectivity_2000}.

For networks, the quality matrix contains more information than expected node growth. As an example, the connection between two layers, measured e.g. by the probability that a random (directional) Markov process along the network would jump from layer A to layer B, is proportional to $\sum_\alpha p_\alpha[1-Q_\alpha]^{-1}_{B,A}m_{\alpha,A}$. Related results were presented in \cite{ferretti_preferential_2011,ferretti_features_2012} for the connectivity as a function of distance or the connectivity between specific classes of nodes. However, these simple results are more involved or impossible for hypergraphs, since there may be complex correlations between multiple nodes sharing the same hyperlink.

The power of our approach lies in its generality. This paper illustrates how heterogeneities may change the degree distribution of models based on preferential attachment, but most such models retain their typical multi-power-law tail. The universality of the degree of the tail for each type of node could be used to test if real-work multi-layer hypergraphs can be well described by preferential attachment dynamics. 

\bibliography{biblio}





\end{document}